\documentclass[pmlr]{jmlr}

\usepackage{longtable}% for long tables

\usepackage{booktabs}
\usepackage[load-configurations=version-1]{siunitx} % newer version
\usepackage{threeparttable}
\usepackage{xcolor}
\usepackage[utf8]{inputenc} % allow utf-8 input
\usepackage[T1]{fontenc}    % use 8-bit T1 fonts
\usepackage{hyperref}       % hyperlinks
\usepackage{url}            % simple URL typesetting
\usepackage{booktabs}       % professional-quality tables
\usepackage{amsfonts}       % blackboard math symbols
\usepackage{nicefrac}       % compact symbols for 1/2, etc.
\usepackage{microtype}      % microtypography
\usepackage{multirow}
\usepackage{graphicx}
\usepackage{amsmath}
\usepackage{array}
\usepackage{threeparttable}
\theorembodyfont{\upshape}
\theoremheaderfont{\scshape}
\theorempostheader{:}
\theoremsep{\newline}

\jmlrvolume{XX}
\jmlryear{2019}
\jmlrworkshop{Machine Learning for Health (ML4H) at NeurIPS 2019}

\title[On the design of CNN for automatic detection of Alzheimer’s disease]{On the design of convolutional neural networks for automatic detection of Alzheimer's disease}

 \author{\Name{Sheng Liu} \Email{shengliu@nyu.edu}\\
 \addr Center for Data Science \\
       New York University
 \AND
 \Name{Chhavi Yadav} \Email{chhavi@nyu.edu}\\
 \addr Courant Institute of Mathematical Sciences \\
       New York University
  \AND
 \Name{Carlos Fernandez-Granda}\thanks{Joint corresponding/last authors.} \Email{cfgranda@cims.nyu.edu}\\
 \addr Center for Data Science\\
 Courant Institute of Mathematical Sciences \\
       New York University
  \AND
 \Name{Narges Razavian}\footnotemark[1]%\thanks{Co-corresponding authors} 
 \Email{narges.razavian@nyumc.org}\\
 \addr  Departments of Population Health and Radiology\\
        Center for Data Science \\
       New York University Langone Medical Center
 }

\begin{document}

\maketitle

\begin{abstract}
Early detection is a crucial goal in the study of Alzheimer's Disease (AD). 
% can impact the chances of success in treatment design, but so far, this task remains challenging. 
In this work, we describe several techniques to boost the performance of 3D convolutional neural networks trained to detect AD using structural brain MRI scans.
Specifically, we provide evidence that (1) instance normalization outperforms batch normalization, (2) early spatial downsampling negatively affects performance, (3) widening the model brings consistent gains while increasing the depth does not, and (4) incorporating age information yields moderate improvement. Together, these insights yield an increment of approximately 14\% in test accuracy over existing models when distinguishing between patients with AD, mild cognitive impairment, and controls in the ADNI dataset. Similar performance is achieved on an independent dataset. We make our code and models
publicly available at \url{https://github.com/NYUMedML/CNN_design_for_AD}. 
\end{abstract}

\section{Introduction}
\label{sec:intro}

Alzheimer's disease (AD) is the leading cause of dementia, and the 6th leading cause of death in the U.S. (\cite{nchs}). Unfortunately, all clinical trials to reverse AD have failed so far (\cite{servick2019another}). It is hypothesized that clinical trials need to target patients at earlier stages before significant brain atrophies. But diagnosing the disease at an early stage is challenging. The current method for early detection relies on PET imaging, which is invasive and very costly. Various studies show that AD-related brain degeneration begins years before the clinical onset of symptoms (\cite{jagust2018imaging}). This suggests that early detection of AD might be possible from standard structural brain imaging scans. Unfortunately, both clinical and also research-grade detection accuracies remain low. 

In this paper we focus on learning to differentiate between cognitively normal aging (CN), mild cognitive impairment (MCI), and Alzheimer's disease (AD), using structural brain MRI (T1-weighted scans). We propose a 3D convolutional neural network (CNN) architecture that achieves state-of-the-art performance for this task. The key novel components of the architecture are
    (1) \textbf{instance normalization}, an alternative to batch normalization introduced originally in the context of style transfer (\cite{ulyanov2016instance,huang2017arbitrary}),
    (2) the use of \textbf{small-sized kernels} in the first layer to avoid downsampling,
    (3) \textbf{wide} architectures with large numbers of filters and relatively few layers, 
    (4) providing the \textbf{age} of the patient to the network through an embedding inspired by a recent technique from natural language processing (\cite{vaswani2017attention}).

Section~\ref{sec:data} describes the data and our preprocessing scheme. Our methodology is then presented in Section~\ref{method}. In Section~\ref{sec:results} we report ablation experiments on the test set to isolate the effect of the different elements in our model, as well as additional analysis of the results. Code to reproduce our main results is publicly available at \url{https://github.com/NYUMedML/CNN_design_for_AD}.

\section{Related Work}
An important task in automatic diagnostics of AD is to distinguish patients with different degrees of mental impairment from MRI scans. Initial works applied simple classifiers such as support vector machines on features obtained from volumetric measurements of the hippocampus (\cite{gerardin2009multidimensional}) and other brain areas (\cite{plant2010automated}).  %One limitation of these studies was the small number of patients (30-40) per category.

More recently, several deep-learning approaches have been applied to this task. \cite{gupta2013natural} used pretraining based on a sparse autoencoder to perform classification on the Alzheimer's Disease Neuroimaging Initiative (ADNI) dataset (\cite{adni}). %This was done to compensate for the small datasets available for training. %previously. 
\cite{hon2017towards} applied state-of-the-art architectures such as VGG~(\cite{simonyan2014very}) and Inception Net (\cite{szegedy2015going}) on the OASIS dataset (\cite{marcus2010open}), selecting the most informative slices in the 3D scans based on image entropy. % They worked with OASIS dataset \cite{marcus2010open} in this study. 
\cite{valliani2017deep}, showed that a ResNet (\cite{he2016deep}) pretrained on ImageNet (\cite{deng2009imagenet}) outperformed a baseline 2D CNN.  \cite{hosseini2016alzheimer} evaluated a 3D CNN architecture on ADNI and data from the CADDementia challenge (\cite{bron2015standardized}). \cite{cheng2017classification} proposed a more computationally-efficient approach based on large 3D patches processed by individual CNNs, which are then combined by an additional CNN to produce the output. \cite{lian2018hierarchical} proposed a related hierarchical CNN architecture that automatically identifies significant patches. Siamese networks were applied by \cite{khvostikov20183d} to distinguish regions of interest  around the hippocampus fusing data from multiple imaging modalities. 
 % However, their data split was found to be leaky.

As described in a recent survey paper, \cite{wen2019convolutional}, many existing works suffer from \emph{data leakage} due to flawed data splits, biased transfer learning, or the absence of an independent test set. 
%They also proposed an open-source framework called Clinica for data processing, which has been used in our study. 
The authors also report that, in the absence of data leakage, CNNs achieve an accuracy of 72-86\% when distinguishing between AD and healthy controls.
In a similar spirit, \cite{fung2019alzheimer} studied the effect of different data-splitting strategies on classification accuracy. % of the three-class classification models.
% They trained popular 2D models pretrained on Imagenet and a variant of 3D ResNet-18. % The splitting strategies considered were: splitting by MRIs randomly ignoring which subject they are from, splitting the data by subject where all the MRI scans from one particular subject were put either in the training or testing set, and splitting the data by visit history. 
A significant drop in test accuracy (from 84\% to 52\% for the three-class classification problem considered in the present work) was reported when there was no patient overlap between the training and test sets. \cite{backstrom2018efficient} also studied the effect of splitting strategies and report similar results for two-way classification.% Additionally, this paper also provided the subject IDs in either sets for reproducibility. The current state of the art classification results without data leakage, for differentiating between CN, MCI and AD is $52.4 \pm 1.8 \% $ (from \cite{fung2019alzheimer}).

\section{Datasets and Preprocessing}
\label{sec:data}
%In this section, we describe the data the setup we used for this study, including the dataset we worked with, and the preprocessing steps we took to produce our results. 
%
\subsection{Datasets} 

%Due to data access and patient privacy issues, open-source medical datasets for Alzheimer’s Disease (AD) detection are often difficult to obtain. High costs of collecting data often result in datasets that are small for deep learning models. As more data often beats sophisticated algorithms, scarcity of data is a bottleneck in training high performing deep learning models for AD detections. ADNI \cite{ADNI Website: \url{http://adni.loni.usc.edu/}, as one of the largest publicly available datasets on Alzheimer's Disease, is an excellent initiative to remove the data availability obstacles from research even though in terms of the size, it is still nowhere close to the computer vision benchmark
%datasets that state of the art deep learning models are trained on. 
For this study, we use T1-weighted structural MRI scans from the ADNI dataset (\cite{mueller2005ways}), which have undergone specific image preprocessing steps including multiplanar reconstruction (MPR), Gradwarp, B1 non-uniformity correction, and N3 intensity normalization (\cite{alzheimer2008adni}). In total, we used over 3000 preprocessed scans. According to the ADNI procedures manuals, labels in the ADNI dataset are extracted based on the scores obtained on memory tasks-- corrected by education level-- and other criteria, some of which are subjective (\cite{alzheimer2008adni}). The labels are AD (mildly demented patients diagnosed with AD), MCI (mildly cognitively-impaired patients in the prodromal phase of AD) and CN (elderly control participants). % These three classes are not exclusive and their inclusion-exclusion criteria can be subjective \cite{ADNI_mannual}. % as it depends on memory complaints from the patients.

\subsection{Data preprocessing}
%As the prominence of extraneous parts of heads captured in MRI scans can potentially bring noise into the model training, multiple preprocessing techniques and packages are adopted in the literature. 
Most previous studies use packages such as \cite{FSL}, Statistical
Parametric Mapping (\cite{SPM}), and FreeSurfer (\cite{fischl2012freesurfer}) to preprocess the data. FSL provides brain extraction and tissue segmentation functionality, while SPM realigns, spatially normalizes, and smooths the scans. FreeSurfer provides a preprocessing stream that includes skull stripping, segmentation, and nonlinear registration. For this study, we used the~\cite{Clinica} software platform developed by \cite{ARAMISLab}, which supports FSL, SPM and FreeSurfer. We first split patients into training, validation and test sets. Then we use Clinica to register the scans to a Dartel template computed exclusively from the training data (\cite{ashburner2007fast}), and normalize them to the Montreal Neurological Institute (MNI) coordinate space (\cite{evans19933d}). The validation and test data are not used to compute any templates in order to avoid data leakage. The input to the Clinica software is the ADNI scans converted to BIDS format. The output dimensions are 121 $\times$ 145 $\times$ 121 voxels along sagittal, coronal and axial dimensions respectively. Due to preprocessing and registration errors, the final number of scans in our dataset is 2702. %The overall dimension of the outputs is still high, restraining the application of recent neural network architectures with a vast amount of parameters. Some works take 2D slices on different views to perform dimensionality reductions, while we used 3D volumes.

% \subsection{Demographics \label{sec:dem}}
  The subjects in the dataset are split between training (70\%), validation (15\%) and test (15\%) sets. As mentioned in the previous section, the split is carried out before preprocessing to avoid any data leakage. Data leakage resulting from using the same subjects in the training and test sets has been shown to artificially improve model performance by a large margin (\cite{backstrom2018efficient,fung2019alzheimer}).  Table~\ref{tab:dem} shows the demographics of the patients in the training, validation, and test sets.
\begin{table}[h]

  \centering
  \resizebox{0.6\linewidth}{!}{
  \begin{tabular}{llllll}
    \toprule
    Split &  
    Class  & Num. subjects     & Num. Scans & Mean Age (std) \\
    \midrule
    \multirow{3}{*}{Train}&CN & 140 & 567 & 77.0~(5.4)\\
    &MCI & 248 &  840 & 75.9~(7.3)\\
     &AD & 193 & 527 & 76.7~(7.4)\\
    \midrule
    \multirow{3}{*}{Val} &  
   CN & 33 & 126 & 77.2~(5.6)\\
    &MCI & 39 & 138& 73.3~(7.2)\\
     &AD & 41 & 124 & 76.1~(8.3)\\
     \midrule
     \multirow{3}{*}{Test} &  
    CN & 24 & 105 & 79.0~(6.1)\\
    &MCI & 43 & 140 & 76.7~(6.5)\\
     &AD & 45 & 135 & 76.4~(5.1)\\
    \bottomrule
  \end{tabular}
  
  }
  \caption{Demographics of our training, validation and test sets after preprocessing.
      %for the output scans of the Clinica pipeline. 
      }
        \label{tab:dem}
\end{table}
\vspace{-1mm}
\section{Methodology \label{method}}
\begin{figure}[t]
\begin{center}
\includegraphics[scale=0.23]{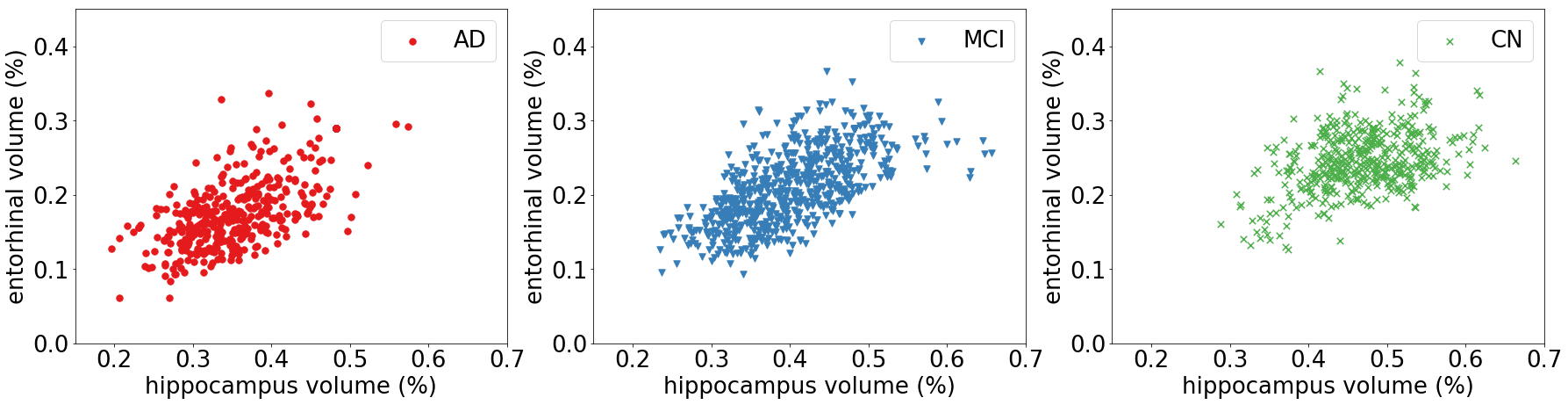}
\end{center}
\vspace*{-5mm}
\caption{Visualization intracranial normalized hippocampus and entorhinal volumes of AD, MCI, and CN subjects. Note that there is significant overlap between the three classes.}
\vspace*{-8mm}
\label{fig:volume_basline}
\end{figure}

Figure~\ref{fig:volume_basline} shows a scatterplot of the values of two popular hand-crafted features associated to AD diagnostics: normalized hippocampus volumes and entorhinal volumes~(\cite{frisoni1999hippocampal,leandrou2018hippocampal}). The features are informative (AD patients tend to have smaller volumes with respect to healthy controls), but they do not enable accurate classification due to the significant overlap between the three classes. This motivates learning discriminative features automatically. Our proposed methodology achieves this using a deep convolutional neural network, inspired by their success in computer vision. However, it is worth emphasizing that our dataset of interest is very different to the datasets of natural images typically used to benchmark computer vision tasks. In our case, all scans are registered and have very similar structure. In addition, the number of examples is usually orders of magnitude smaller. Therefore, we need to design architectures capable of learning subtle differences from relatively small datasets.
\subsection{Proposed model}
%Several recent studies of AD detection \cite{valliani2017deep, abrol2019deep, maqsood2019transfer, yang2018visual} apply three classical architectures to perform classification: AlexNet \cite{krizhevsky2012imagenet}, ResNet \cite{he2016deep}, and VGGNet. VGGNet is structurally close to AlexNet as it does not have skip-connections and uses fully-connected layers. 
Our proposed architecture is a 3D CNN model, composed of convolutional, normalization, activation and max-pooling layers. The architecture is described in more detail in Table~\ref{tab:arch}. In this section, we outline several design choices that significantly boost the performance of the network for the task of differentiating between CN, AD, and MCI patients.

\textbf{Instance normalization.} Batch normalization, introduced by \cite{ioffe2015batch}, has become one of the standard techniques to ease training of deep feed-forward networks. In our proposed model, however, we apply instance normalization, a technique introduced in the context of style transfer~(\cite{ulyanov2016instance,huang2017arbitrary}). %Instance normalization consists of centering and normalizing feature maps of every individual channel for each input image. Intuitively, this may render the network features more invariant to systematic differences between scans. 
In Section \ref{sec:norm}, we show that applying instance normalization consistently outperforms batch normalization for our task of interest. 

\textbf{Small-sized kernels.} In contrast to most standard architectures for image classification, we use small-sized kernels in the first convolutional layer to prevent early spatial downsampling. For instance, ResNet and AlexNet use relatively large kernel sizes and strides in their first layer, which dramatically reduce the spatial dimension of their inputs. This accelerates the computation and is not usually detrimental in the case of natural-image classification tasks. However, for our task of interest, early downsampling results in significant loss of performance, as we show in Section
\ref{sec:kernel}.

\textbf{Wider network.} In our architecture design we favor a wider architecture that is not too deep. In Section~\ref{sec:wide}, we find that increasing the depth of the model only brings marginal gains, whereas widening the architecture improves performance significantly. %The best-proposed model uses a widening factor of 8.

\textbf{Age encoding.} Brains typically shrink to some degree in healthy aging (\cite{peters2006ageing}). This might confuse the model since Alzheimer’s disease may have a similar effect (\cite{van1991entorhinal}). A simple way to incorporate age in our model is to concatenate the normalized age of the patient to the output of the convolutional layers. However, this seems to result in worse performance. In order to better integrate age information, we encode each age value into a vector and combine the vector with the output of the convolutional layers. See Appendix~\ref{sec:age_append} for further details.
%\subsection{Adversarial Training}
%As we mentioned in section \ref{sec:dem}, higher accuracy can be obtained when the same subject shows up in both training and test dataset. This is because transitions in disease stage occur at a
%very low frequency in the data set. Models can easily achieve a relatively high performance by memorizing the label for each subject, recognizing the patient and predicting the same label while evaluation. Therefore, a subject-invariant representation should be learned to address this issue. We consider unwanted subject information to be those could help models to recognize which subject the scan comes from. In order to disentangle the unwanted information, we train an adversarial network to actually classify which subject the scan comes from such that the learned representation hard to perform this scan-subject classification task. The details are in the Appendix. 

\begin{table}[t]
  \centering
\resizebox{0.7\linewidth}{!}{
  \begin{tabular}{llll}
    \toprule
    Block & Layer  & Type & Output size \\
    \midrule
    & Inputs &         & $96\times 96\times 96$ \\
     \midrule
    \multirow{4}{*}{1} & Conv3D &   k1-c4$\cdot f$-p0-s1-d1      &  $96\times 96\times 96$\\
    & InstanceNorm3D &       &  \\
    & ReLU     &     &     \\
    & MaxPool3D &  k3-s2  &   $47\times 47\times 47$\\
    \midrule
   \multirow{4}{*}{2} & Conv3D &   k3-c32$\cdot f$-p0-s1-d2      & $43\times 43\times 43$ \\
    & InstanceNorm3D &       &  \\
    & ReLU     &     &     \\
    & MaxPool3D &  k3-s2  & $21\times 21\times 21$  \\
    \midrule
    \multirow{4}{*}{3} & Conv3D &   k5-c64$\cdot f$-p2-s1-d2      & $17\times 17\times 17$ \\
    & InstanceNorm3D &       &  \\
    & ReLU     &     &     \\
    & MaxPool3D &  k3-s2  &  $8\times 8\times 8$ \\
    \midrule
    \multirow{4}{*}{4} & Conv3D &   k3-c64$\cdot f$-p1-s1-d2      & $6\times 6\times 6$ \\
    & InstanceNorm3D &       &  \\
    & ReLU     &     &     \\
    & MaxPool3D &  k5-s2  &  $5\times 5\times 5$ \\
    \midrule
    FC1 &         & 1024 \\
    FC2 & & 3\\
    Softmax & & 3\\
    \bottomrule
    \vspace{-1mm}
  \end{tabular}
  }
  \vspace*{-5mm}
    \caption{The backbone architecture. k = kernel size, c = number of channels as a multiple of the widening factor $f$, p = padding size, s = stride and d = dilation. We report results for $f$ equal to 1, 2, 4, and 8 in Section~\ref{sec:results}. The age encoding, if used, is forward propagated through two linear layers with layer normalization before being added to the output of FC1, see Table \ref{tab:age_arch} in the Appendix for details. }
  \label{tab:arch}
\vspace*{-1mm}
\end{table}
\section{Experiments and Results \label{sec:results}}
In this section, we present and interpret the results of our study, which demonstrate the effectiveness of the techniques described in Section \ref{method}.
\begin{table}[t]
  \centering
   \resizebox{0.9\linewidth}{!}{
  \begin{threeparttable}
  \begin{tabular}{lllll}
    \toprule
    Method  & Accuracy     & Balanced Acc & Micro-AUC & Macro-AUC \\
    \midrule
    ResNet-18\tnote{$\star$} &  $50.8\%$      & - & - & - \\
    ResNet-18 pretrained\tnote{$\star$} &  $56.8\%$       & - & - & -\\
    ResNet-18 3D\tnote{$\diamond$} &  $ 52.4 \pm 1.8\%$       & $ 53.1\%$  & - & -\\
    ResNet-18 3D    &  $50.1 \pm 1.1\%$       & $51.3 \pm 1.0\%$ & $71.2 \pm 0.4\%$ & $72.4 \pm 0.7\%$ \\
    AlexNet 3D &  $57.2 \pm 0.5\%$       & $56.2 \pm 0.8\%$ & $75.1 \pm 0.4\%$ & $74.2 \pm 0.5\%$ \\
    %proposed\tnote{$\bullet$}  on AIBL     & $63.6 \pm 0.7\%$      & $65.7 \pm 1.1\%$ & $90.0 \pm 0.6\%$ & $82.1 \pm 0.7\%$\\
    proposed\tnote{$\bullet$}    & $66.9 \pm 1.2\%$      & $67.9 \pm 1.1\%$ & $82.0 \pm 0.7\%$ & $78.5 \pm 0.7\%$\\
%    proposed + AT & $67.1 \pm 1.1\%$      & $68.0 \pm 0..9\%$ & $79.2 \pm 0.7\%$ & $79.4 \pm 0.8\%$\\
        %proposed\tnote{$\bullet$} ~+ Age (direct) & $61.5\pm 1.4\%$      & $62.6\pm 1.0\%$ & $78.6  \pm 1.2\%$ & $78.3 \pm 1.1\%$\\
    proposed\tnote{$\bullet$} ~+ Age & $\bf{68.2\pm 1.1}\%$      & $\bf{70.0\pm 0.8\%}$ & $\bf{82.0  \pm 0.2\%} $ & $\bf{80.0 \pm 0.5\%}$\\

%    proposed + Age + AT & $\bf{68.4 \pm 0.8\%}$      & $69.6 \pm 0.8\%$ & $\bf{82.9 \pm 0.6\%}$ & $\bf{81.5 \pm 0.3\%}$\\
    \bottomrule
  \end{tabular}
 
\begin{tablenotes}\footnotesize
\item[$\star$] \textit{Results on $2D$ ResNets initialized with or without pretrained weights on Imagenet reported by \cite{valliani2017deep}. }
\item[$\diamond$] \textit{$3D$ ResNet with mild modifications, see~\cite{fung2019alzheimer} for details. The balanced accuracy is computed using the confusion matrix in the paper.}
\item[$\bullet$] \textit{The backbone model showed in Table~\ref{tab:arch} with a widening factor of $8$.}
\end{tablenotes}
\end{threeparttable}
 }
    \caption{Comparison of the published models to our best proposed models. + Age means that the model incorporates age encodings.} %{and AT means with adversary training. }
  \label{tab: result_comp}
  \vspace*{-6mm}
\end{table}
\begin{figure}[t]
\begin{center}
\includegraphics[scale=0.4]{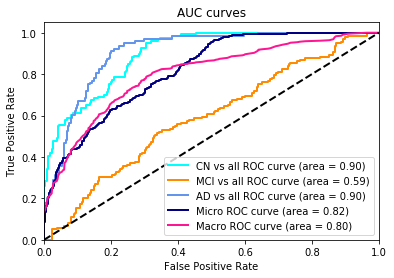}
\includegraphics[scale=0.4]{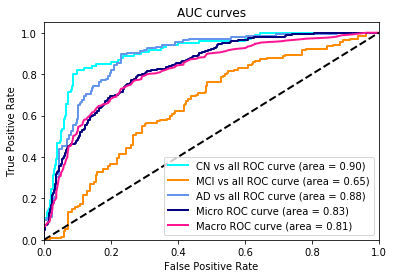}
\end{center}
\vspace*{-5mm}
\caption{ROC curves on the validation set (left) and test set (right). Differentiating CN or AD from all other classes results in high AUCs while detecting MCI remains a difficult task.}
\vspace*{-5mm}
\label{fig:rocs}
\end{figure}
\vspace*{-3mm}
% \subsection*{Ad
\subsection{Description of computational experiments}
We choose AlexNet and ResNet as baseline 3D CNNs since they are popular in computer vision as well as for our task. Unsurprisingly, given the size of the dataset, all architectures, including ResNet and AlexNet, are able to fit the training set with high balanced accuracy, while the generalization ability varies. We perform data augmentation via Gaussian blurring with $\sigma$ uniformly chosen from 0 to 1.5, and random cropping of size $96 \times 96 \times 96$. We set the batch size to 4 (for memory considerations) and the learning rate to 0.01. We use stochastic gradient descent with momentum equal to 0.9. We use the same settings for AlexNet and ResNet, except for the batch size which is set to 16 since these architectures use batch normalization. After training, the models with the lowest validation loss are saved and evaluated on the test set to obtain the results reported in Table \ref{tab: result_comp}. We compute the confidence intervals using bootstrapping.

\subsection{Comparison to other methods}
Our primary metric in this work is standard classification accuracy (Acc). As the test set is not necessarily balanced, we also use balanced classification accuracy (Bal-Acc) which is calculated as the average of the recall of each class. We also compute area under the ROC curves (AUCs), which are widely used for measuring the predictive accuracy of binary classification problems. This metric indicates the relationship between the true positive rate and false positive rate when the classification threshold varies. As AUC can only be computed for binary classification, we compute AUCs for all three binary problems of distinguishing between one of the categories and the rest. We also calculate micro and macro averages, denoted as Micro-AUC and Macro-AUC respectively. 

Table \ref{tab: result_comp} summarizes our results. Our proposed model significantly outperforms previously reported results\footnote{Some of the results in the literature use different data splits. However, we also report 3D the two most popular models (ResNet-18 and AlexNet) trained using the same split as the proposed model.}, as well as the baseline architectures. Incorporating age through the proposed encoding improves performance moderately.  We show the ROC curves obtained on the validation and test set in Figure \ref{fig:rocs}. The model achieves around $90\%$ AUC when distinguishing CN or AD from the other two classes, and $60-65\%$ when distinguishing MCI from the other two classes.
\vspace*{-3mm}
\subsection{Ablation studies}
In this section, we perform ablation studies on the techniques described in Section~\ref{method} to isolate their individual contributions to the accuracy of the proposed model. The studies were performed on the test set.

\subsubsection{\textbf{Instance normalization vs batch normalization} \label{sec:norm}}
We compare batch normalization (BN) and instance normalization (IN) on the backbone architecture using different widening factors and on ResNet-18. The results are in Table~\ref{tab:normalization}. More comprehensive evaluations on different widening factors are presented in  Table~\ref{tab:normalization_app} of Appendix~\ref{sec: invsbn}. Models with IN layers perform consistently better than models with BN layers.  % We hypothesize that this is because the MRI contrast as well as some non-localized noise introduced by machine hardware can be normalized using IN layers.  
\begin{table}[h]
  \centering
\resizebox{0.9\linewidth}{!}{
  \begin{tabular}{lllll}
    \toprule
    Method  & Accuracy     & balanced Acc & Micro-AUC & Macro-AUC \\
    \midrule
   %$\times 1$ with IN & $\bf{56.4 \pm 1.4\%}$      & $\bf{54.8 \pm 1.2\%}$ & $\bf{74.2 \pm 0.8\%}$ & $\bf{75.6  \pm 0.9\%}$\\
    %$\times 1$ with BN & $54.2 \pm 1.2\%$       & $53.3 \pm 0.8\%$ & $74.1 \pm 0.7\%$ & $73.2  \pm 0.9\%$\\
    
    %$\times 2$ with IN & $\bf{58.4 \pm 1.7\%}$       & $\bf{57.8 \pm 1.7\%}$ & $\bf{77.2 \pm 0.8\%}$ & $\bf{76.6  \pm 0.9\%}$\\
    %$\times 2$ with BN & $57.1 \pm 0.7\%$       & $55.6 \pm 0.8\%$ & $74.8 \pm 0.6\%$ & $73.6  \pm 0.6\%$\\
    %$\times 4$ with IN & $\bf{63.2 \pm 1.0\%}$      & $\bf{63.3 \pm 0.9\%}$ & $\bf{80.5 \pm 0.5\%}$ & $\bf{77.0 \pm 0.7\%}$  \\
    %$\times 4$ with BN & $61.8 \pm 1.1\%$      & $62.2 \pm 1.1\%$ & $77.0 \pm 0.5\%$ & $73.0 \pm 0.6\%$  \\
    $\times 4$ with IN & $\bf{63.2 \pm 1.0\%}$      & $\bf{63.3 \pm 0.9\%}$ & $\bf{80.5 \pm 0.5\%}$ & $\bf{77.0 \pm 0.7\%}$  \\
    $\times 4$ with BN & $61.8 \pm 1.1\%$      & $62.2 \pm 1.1\%$ & $77.0 \pm 0.5\%$ & $73.0 \pm 0.6\%$  \\
     $\times 8$ with IN & $\bf{66.9 \pm 1.2\%}$      & $\bf{67.9 \pm 1.1\%}$ & $\bf{82.0 \pm 0.7\%}$ & $\bf{78.5 \pm 0.7\%}$  \\
    $\times 8$ with BN & $58.8 \pm 0.9\%$      & $60.7 \pm 0.7\%$ & $75.9 \pm 0.7\%$ & $73.1 \pm 0.8\%$\\
    ResNet-18 with IN & $\bf{52.3 \pm 0.8\%}$      & $\bf{52.7 \pm 1.1\%}$ & $\bf{74.1 \pm 0.7\%}$ & $\bf{73.1 \pm 0.9\%}$\\
    ResNet-18 with BN & $50.1 \pm 1.1\%$       & $51.3 \pm 1.0\%$ & $71.2 \pm 0.4\%$ & $72.4 \pm 0.7\%$\\
    \bottomrule
  \end{tabular}
  }
  \vspace*{1mm}
    \caption{Comparison of batch normalization (BN) and instance normalization (IN) layers on the backbone architecture with widening factor of 4 and 8 and on ResNet-18. See also Table~\ref{tab:normalization_app} for results with other widening factors. Instance normalization outperforms batch normalization in all cases.}
  \label{tab:normalization}

\end{table}

\subsubsection{\textbf{Early spatial downsampling} \label{sec:kernel}}
\begin{figure}[h]
\begin{center}
\includegraphics[scale=0.4]{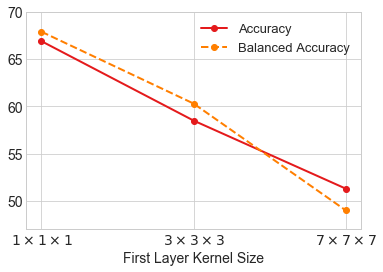}
\includegraphics[scale=0.4]{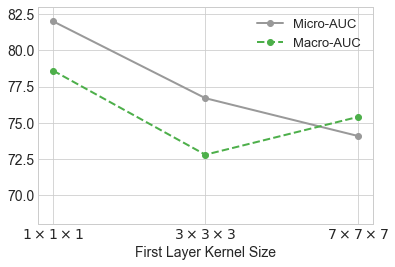}
\end{center}

\caption{Comparison of the performances of different first layer kernel sizes for the backbone architecture in Table \ref{tab:arch}. Larger kernel sizes in the first layer result in worse performance.}

\label{fig:first_kernel}
\end{figure}

Here we study how the kernel size of the first convolutional layer affects the final classification performance. We compare original kernel sizes $1\times1\times1$ with stride 1, $3 \times 3 \times 3$ with stride 2, and $7 \times 7 \times 7$ with stride 4. The results are summarized in Figure \ref{fig:first_kernel}. The smallest kernel has the best performance. This is a possible explanation for the inferior performance of ResNet and AlexNet for our task. We further check this hypothesis for ResNet, the results show that reducing kernel size for the initial layer is effective for ResNet as well (see details in Appendix~\ref{sec:first_kernel_append}).

%\newpage
\subsubsection{\textbf{Wider or deeper model?} \label{sec:wide}}
\begin{figure}[h]
\hspace{-0.6cm} 
\centering
\begin{tabular}{ >{\centering\arraybackslash}m{0.31\linewidth} >{\centering\arraybackslash}m{0.31\linewidth}}
%\textbf{Raw Data}} & \multicolumn{2}{c}{\textbf{First 3 NMF-TS coefficients}}\\
Width & Depth\\
 \includegraphics[width=\linewidth]{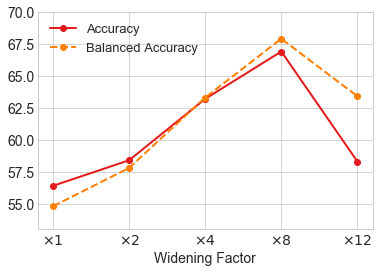}  & \includegraphics[width=\linewidth]{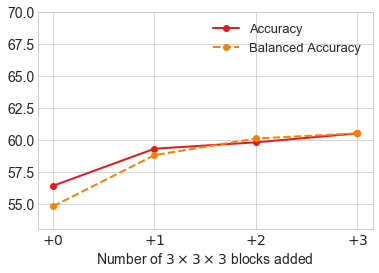} \\ \includegraphics[width=\linewidth]{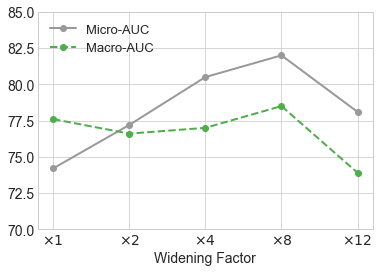}& \includegraphics[width=\linewidth]{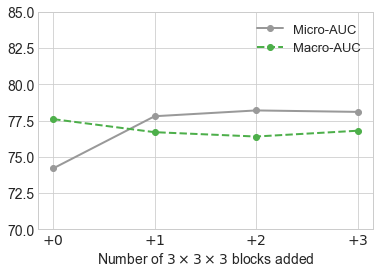} 
 \end{tabular}
 \caption{Performance for different widening factors (left) and numbers of added blocks (right) for backbone architecture in Table \ref{tab:arch}. Wider architectures consistently achieve better performance up until a widening factor of x4. Deeper networks only achieve marginal improvement.}
\label{fig:width}
\end{figure}

In this section we compare the effect of varying width or depth on classification accuracy. 
The left graph in Figure \ref{fig:width} shows that widening the network architecture leads to better classification performance up until a certain point. This finding is in line with results reported for the ResNet by \cite{zagoruyko2016wide}. We increase the depth of our backbone network by adding convolutional blocks (convolutional layers + instance normalization + ReLU activation). It should be noted that the size of the representation output from the final convolutional block might decrease when the network becomes deeper. To control for the effects of the representation size when making the architecture deeper, convolutional layers in each block are set to have kernel size of $3\times 3 \times 3$, stride of 1 and padding of 1. Increasing depth only achieves small gains in accuracy. We also observe that deeper networks are often slower and more difficult to train when compared to wider networks. 
\vspace*{-2mm}
\subsubsection{\textbf{Impact of dataset size}}
\begin{figure}[h]
\begin{center}
\includegraphics[scale=0.45]{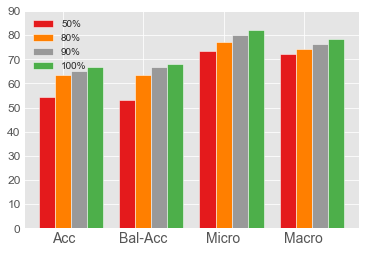}
\end{center}
\vspace*{-6mm}
\caption{Performance of the proposed model evaluated using different subsampling rates. The trend is clear: increasing dataset size improves performance across all evaluation metrics.}
\label{fig:perc_data}

\end{figure}

%In principle, the effectiveness of the proposed model can be improved with a larger dataset. 
In Figure~\ref{fig:perc_data}, we report the performance of the proposed model for datasets of different sizes (obtained by randomly subsampling the data). 
We observe that increasing the size of the dataset results in better performance in all evaluation metrics. Given that the model is trained on a very small dataset compared to regular computer-vision tasks, more data may be needed to exhaust the representation ability of the models.

\subsection{Validation with independent dataset}

We test the generalization capacity of our model on  a completely separate dataset, obtained from the Australian Imaging Biomarkers and Lifestyle flagship study of ageing (AIBL)~(\cite{ellis2009australian}). We follow the same preprocessing procedures as for the ADNI validation and test set (described in Section~\ref{sec:data}), being careful to avoid any data leakage. After preprocessing, we obtain 783 CN scans from 461 subjects with average age 73.5, 150 MCI scans from 113 subjects with average age 76.2, and 134 AD scans from 95 subjects with average age 75.4. The results are shown in Table~\ref{tab:aibl}. We apply our proposed architecture without age information, since this information may not be readily available for different datasets. The model achieves a similar performance on this independent dataset as on the ADNI data, which demonstrates that the features learned by the network generalize effectively.

\begin{table}[h]
  \centering
  \resizebox{0.8\linewidth}{!}{
  \begin{tabular}{lllll}
    \toprule
    Method  & Accuracy     & Balanced Acc & Micro-AUC & Macro-AUC \\
    \midrule
    
    proposed on ADNI     & $66.9 \pm 1.2\%$      & $67.9 \pm 1.1\%$ & $82.0 \pm 0.7\%$ & $78.5 \pm 0.7\%$\\
    proposed on AIBL     & $63.6 \pm 0.7\%$      & $65.7 \pm 1.1\%$ & $90.0 \pm 0.6\%$ & $82.1 \pm 0.7\%$\\
%    proposed + AT & $67.1 \pm 1.1\%$      & $68.0 \pm 0..9\%$ & $79.2 \pm 0.7\%$ & $79.4 \pm 0.8\%$\\
        %proposed\tnote{$\bullet$} ~+ Age (direct) & $61.5\pm 1.4\%$      & $62.6\pm 1.0\%$ & $78.6  \pm 1.2\%$ & $78.3 \pm 1.1\%$\\

%    proposed + Age + AT & $\bf{68.4 \pm 0.8\%}$      & $69.6 \pm 0.8\%$ & $\bf{82.9 \pm 0.6\%}$ & $\bf{81.5 \pm 0.3\%}$\\
    \bottomrule
  \end{tabular}
  }
\vspace*{-2mm}
    \caption{Comparison of the performance of the proposed model on the ADNI and AIBL datasets.} %{and AT means with adversary training. }
  \label{tab:aibl}
\end{table}
\vspace*{-6mm}
\section{Analysis}
\subsection{Analysis of wrongly-classified subjects}
\begin{figure}[h]
\vspace*{-3mm}
\begin{center}
\includegraphics[scale=0.47]{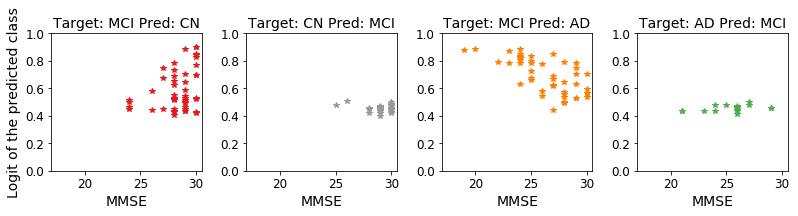}

\end{center}
\vspace*{-7mm}
\caption{The Mini-Mental State Exam (MMSE) scores and corresponding logits of the predicted class for wrongly-classified validation examples.}
\label{fig:class_conf}
\vspace*{-8mm}
\end{figure}

We analyze the wrongly-classified validation examples in Figure~\ref{fig:class_conf}. Mini-Mental State Exam (MMSE) scores (with value ranges from 0 to 30) are widely used tools for detecting cognitive impairment, assessing severity, and monitoring cognitive changes over time. Lower scores often mean more cognitive impairment. The model's output after the softmax layer (logits) can be viewed as the confidence of the model in predicting a class. The trend in the figure shows that for higher MMSE scores the model becomes more confident in predicting CN, and less confident in predicting AD. Since the criteria to assign labels are subjective, and the boundary between MCI and the other two classes is not always clear, it is possible that some of the classification errors are due to noise in the labels. % The trends in the figure can indicate that our learned model is capable of finding the mapping between input scan and its cognitive impairment level. 

\subsection{Opening the black box}
\begin{figure}[h]
\begin{center}
\begin{tabular}{
>{\centering\arraybackslash}m{0.03\linewidth}
>{\centering\arraybackslash}m{0.18\linewidth}
>{\centering\arraybackslash}m{0.18\linewidth}
>{\centering\arraybackslash}m{0.18\linewidth} >{\centering\arraybackslash}m{0.18\linewidth}}
%\textbf{Raw Data}} & \multicolumn{2}{c}{\textbf{First 3 NMF-TS coefficients}}\\
& Axial 50th  & Axial 26th & Coronal 56th & Sagittal 26th \\
Agg. & \includegraphics[scale=0.33]{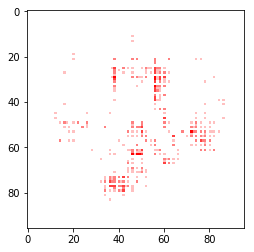} & \includegraphics[scale=0.33]{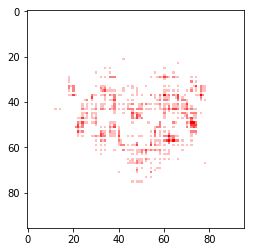}& \includegraphics[scale=0.33]{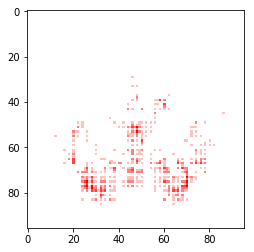}&\includegraphics[scale=0.33]{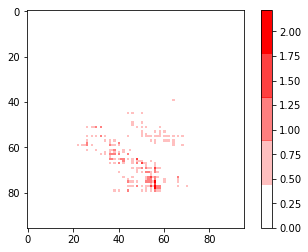}\\
CN & \includegraphics[scale=0.33]{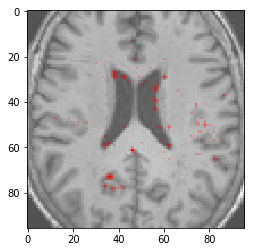} & \includegraphics[scale=0.33]{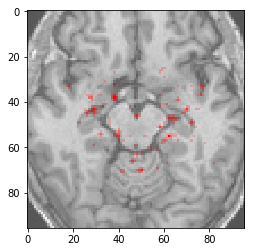}& \includegraphics[scale=0.33]{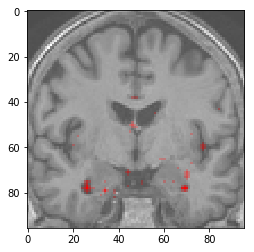}&\includegraphics[scale=0.33]{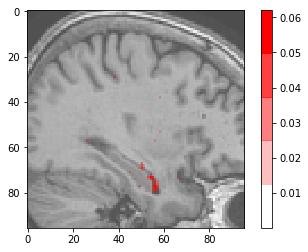}\\
MCI & \includegraphics[scale=0.33]{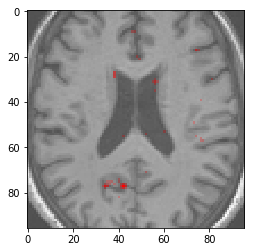} & \includegraphics[scale=0.33]{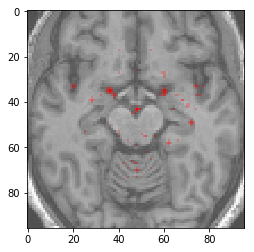}& \includegraphics[scale=0.33]{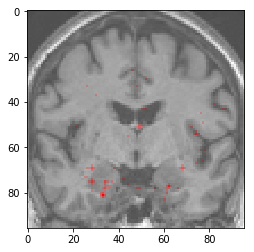}&\includegraphics[scale=0.33]{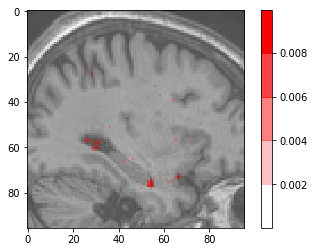}\\
AD & \includegraphics[scale=0.33]{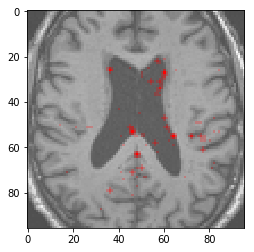} & \includegraphics[scale=0.33]{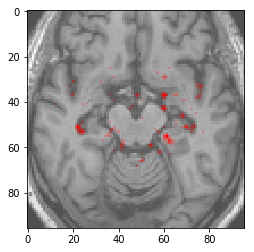}& \includegraphics[scale=0.33]{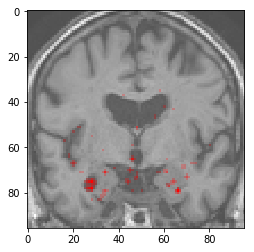}&\includegraphics[scale=0.33]{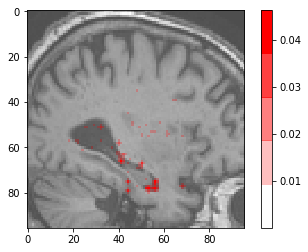}
 \end{tabular}

\end{center}
\vspace*{-7mm}
\caption{Visualization of class saliency maps (slices) obtained by computing the magnitude of the gradient of the learned map associated to each class with respect to the input (for each patient we compute the gradient with respect to their true class). The top row shows a aggregated plot of all saliency maps in the validation set for three slices. The bottom rows show saliency maps for examples of patients in each class superposed on the corresponding registered brain scan and smoothed by a Gaussian kernel with $\sigma=0.8$. %Higher intensity of red color indicates that the region is more significant for classification.
}
\label{fig:visualziation}
\end{figure}

%In this section, we provide a deeper analysis of what information our proposed model extracts from an MRI scan to estimate the cognitive impairment level. 
In order to visualize the features learned by the model, we compute saliency maps consisting of the magnitude of the gradient of the target class score with respect to the input (\cite{simonyan2013deep}). Figure~\ref{fig:visualziation} shows examples of these saliency maps for randomly selected scans in the validation set belonging to each class. It also shows aggregated maps that combine saliency maps from all scans in the validation set. These results reveal some interesting aspects of the proposed model: the model focuses on gray-matter regions around the hippocampus and the ventricles, which is consistent with existing biomarkers~(\cite{risacher2013neuroimaging}), as well as on some additional regions. A detailed study of these regions lies beyond the scope of this work, but is an intriguing direction for future research.   

\vspace*{-8mm}
\section{Conclusion}
In this paper, we develop a novel 3D CNN architecture to perform three-way classification between patients with Alzheimer's disease, patients with mild cognitive impairment, and healthy controls. Our architecture combines different elements (instance normalization, wider layers, and an encoding of the patient's age) to achieve a significant gain in classification accuracy, demonstrated on completely held-out data and on an independent dataset.
% The proposed architecture uses instance normalization rather than batch normalization. We show that larger-sized kernels in the initial layer degrades performance. We also demonstrate that widening a simple CNN model can boost performance while increasing depth does not. Combining these insights yields a significant gain in classification accuracy, which is tested on completely held-out data and on an independen dataset.

%The proposed CNN also opens up several possibilities for future exploration. First, is it possible to make similar architecture work on other types of inputs such as segmentations and PET scans? Second, can similar methods work on other types of supervised learning tasks, especially to other tasks which require learning subtle differences among images? Although our discussion of these directions is still speculative, we are excited about the possibilities it opens up and hope our observations will prove useful for future development.

%While we boost the performance by a large margin with the proposed model, there are still several difficulties in this task. Firstly, the inclusion and exclusion criteria for the diagnosis make them noisy however a more accurate diagnosis is hard to obtain. 

%TODO: Add a paragraph on what are the implications of this model.
\newpage

\acks{
The authors would like to thank Henry Rusinek and 
Arjun Masurkar for their useful comments on earlier versions of this
manuscript. Authors also acknowledge Leon Lowenstein Foundation for funding support, and the Alzheimer's Disease
Neuroimaging Initiative (ADNI) (National Institutes of Health Grant U01 AG024904) and
DOD ADNI (Department of Defense award number W81XWH-12-2-0012), as well as National Institute on Aging and the National Institute of Biomedical Imaging and
Bioengineering for data collection and sharing.}

\bibliography{jmlr-sample}

\newpage
\appendix
%\section*{Appendix}
\section{Age encodings \label{sec:age_append}}
To compute the age encoding vector, we first fix the age values range from 0 to 120 years old, and round all possible age values to 0.5 decimal places. In total, we get 240 possible age values. Inspired by the positional encoding in the transformer model~(\cite{vaswani2017attention}), we use sinusoidal functions to implement the encoding. We define $\text{AE}_{(\text{age})} \in \mathbf{R}^{d_{\text{model}}}$ to be the age encoding function defined as:
\begin{align*}
    \text{AE}_{(\text{age},2i)} &= \sin(\text{age}/10000^{2i/d_{\text{model}}})\\
    \text{AE}_{(\text{age},2i+1)} &= \cos(\text{age}/10000^{2i/d_{\text{model}}})
\end{align*}
where age is one of the 240 possible age values,and $i = 0,1,2,\dots,d_{\text{model}}/2-1$ is the dimension and $d_{\text{model}}$ is the size of the encodings. We further transform the age encodings using a few fully connected layers to match the scales and sizes with the visual representation. The architecture for the transformation is showed in Table \ref{tab:age_arch}. \\

\begin{table}[!h]
  \centering
  \begin{tabular}{lll}
    \toprule
    Layer   & Output size \\
    \midrule
   
     Linear &       512 \\
     LayerNorm  & 512\\
     Linear &     1024\\
    \bottomrule
  \end{tabular}
    \caption{The age encoder architecture}
  \label{tab:age_arch}
\end{table}
We compare this method with a simple baseline. In the baseline, we directly concatenate normalized age (with range from 0 to 1) to the learned representation obtained from the convolutional layers. The results are in Table~\ref{tab:age_comp}. Our proposed encoding results in improved performance, whereas the baseline encoding results in worse performance.
\begin{table}[h]
  \centering
  \begin{threeparttable}
  \begin{tabular}{lllll}
    \toprule
    Method  & Accuracy     & Balanced Acc & Micro-AUC & Macro-AUC \\
    \midrule

No age information    & $66.9 \pm 1.2\%$      & $67.9 \pm 1.1\%$ & $82.0 \pm 0.7\%$ & $78.5 \pm 0.7\%$\\

    Proposed age encoding & $\bf{68.2\pm 1.1}\%$      & $\bf{70.0\pm 0.8\%}$ & $\bf{82.0  \pm 0.2\%} $ & $\bf{80.0 \pm 0.5\%}$\\
    
Baseline age encoding & $61.5\pm 1.4\%$      & $62.6\pm 1.0\%$ & $78.6  \pm 1.2\%$ & $78.3 \pm 1.1\%$\\
    
    \bottomrule
  \end{tabular}
\end{threeparttable}

\vspace*{2mm}
    \caption{Comparison of different ways of incorporating the age information using the proposed architecture.} %{and AT means with adversary training. }
    \label{tab:age_comp}
\end{table}

\section{Instance Normalization vs Batch Normalization \label{sec: invsbn}}
In Table~\ref{tab:normalization_app}, we present the complete results of comparing Instance Normalization (IN) and Batch Normalization (BN) on our backbone architecture with various widening factors. IN consistently outperforms BN for all architectures.
\begin{table}[h]
  \centering
  \begin{tabular}{lllll}
    \toprule
    Method  & Accuracy     & balanced Acc & Micro-AUC & Macro-AUC \\
    \midrule
   $\times 1$ with IN & $\bf{56.4 \pm 1.4\%}$      & $\bf{54.8 \pm 1.2\%}$ & $\bf{74.2 \pm 0.8\%}$ & $\bf{75.6  \pm 0.9\%}$\\
    $\times 1$ with BN & $54.2 \pm 1.2\%$       & $53.3 \pm 0.8\%$ & $74.1 \pm 0.7\%$ & $73.2  \pm 0.9\%$\\
    
    $\times 2$ with IN & $\bf{58.4 \pm 1.7\%}$       & $\bf{57.8 \pm 1.7\%}$ & $\bf{77.2 \pm 0.8\%}$ & $\bf{76.6  \pm 0.9\%}$\\
    $\times 2$ with BN & $57.1 \pm 0.7\%$       & $55.6 \pm 0.8\%$ & $74.8 \pm 0.6\%$ & $73.6  \pm 0.6\%$\\
    $\times 4$ with IN & $\bf{63.2 \pm 1.0\%}$      & $\bf{63.3 \pm 0.9\%}$ & $\bf{80.5 \pm 0.5\%}$ & $\bf{77.0 \pm 0.7\%}$  \\
    $\times 4$ with BN & $61.8 \pm 1.1\%$      & $62.2 \pm 1.1\%$ & $77.0 \pm 0.5\%$ & $73.0 \pm 0.6\%$  \\
     $\times 8$ with IN & $\bf{66.9 \pm 1.2\%}$      & $\bf{67.9 \pm 1.1\%}$ & $\bf{82.0 \pm 0.7\%}$ & $\bf{78.5 \pm 0.7\%}$  \\
    $\times 8$ with BN & $58.8 \pm 0.9\%$      & $60.7 \pm 0.7\%$ & $75.9 \pm 0.7\%$ & $73.1 \pm 0.8\%$\\
    ResNet-18 with IN & $\bf{52.3 \pm 0.8\%}$      & $\bf{52.7 \pm 1.1\%}$ & $\bf{74.1 \pm 0.7\%}$ & $\bf{73.1 \pm 0.9\%}$\\
    ResNet-18 with BN & $50.1 \pm 1.1\%$       & $51.3 \pm 1.0\%$ & $71.2 \pm 0.4\%$ & $72.4 \pm 0.7\%$\\
    \bottomrule
  \end{tabular}
    \caption{Comparison of batch normalization (BN) and instance normalization (IN) layers on the backbone architecture with different widening factors as well as on ResNet-18, instance normalization outperforms batch normalization in all cases.}
  \label{tab:normalization_app}
\end{table}
\section{Early spatial downsampling \label{sec:first_kernel_append}}

\begin{figure}[h]
\begin{center}
\includegraphics[scale=0.45]{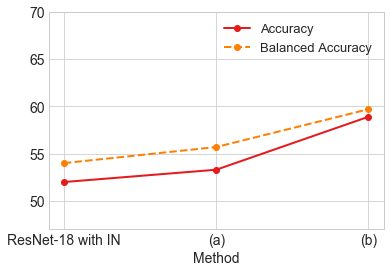}
\includegraphics[scale=0.45]{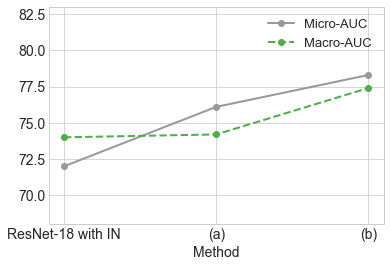}
\end{center}
%\vspace*{-5mm}
\caption{Performance of different first layer kernel sizes for a ResNet-18 with IN. Method $\mathbf{(a)}$ modifying the kernel into $3\times 3 \times 3$, stride 1.  In method $\mathbf{(b)}$, we further add a $1\times 1\times 1$ convolutional block on the top of the model from method $\mathbf{(a)}$.}
\label{fig:first_kernel_append}
\end{figure}

Figure~\ref{fig:first_kernel_append} shows results for ResNet when for different kernel sizes of the first convolutional layer. We modify the architecture of a ResNet-18 with instance normalization in the following way: \textbf{(a)} we reduce the size of the first convolution from $7\times 7 \times 7$ with stride 2 into $3\times3\times3$ with stride 1, \textbf{(b)} we further add a $1\times 1 \times 1$ convolutional block on the top (right after the input), the results are showed in Figure \ref{fig:first_kernel_append}. These results demonstrate that reducing filter size in the first convolutional layer yields performance improvements for the ResNet as well. For a ResNet-18 with batch normalization, performance also improves, although less markedly.

\end{document}